\documentclass[prl,groupedaddress,superscriptaddress,twocolumn]{revtex4-1}

\usepackage[T1]{fontenc}
\usepackage[utf8]{inputenc}
\usepackage{lmodern}
\usepackage{graphicx}
\usepackage{amsmath}
\usepackage{amssymb}
\usepackage{amsfonts}
\usepackage{xcolor}
\usepackage[colorlinks=true, linkcolor=black, hyperindex=true]{hyperref}
\usepackage{epstopdf}

\begin{document}
\title{Multiferroic Two-Dimensional Materials}

\author{L. Seixas}
  \email{seixasle@gmail.com}
  \affiliation{Centre for Advanced 2D Materials and Graphene Research Centre, National University of Singapore, Singapore 117542, Singapore}

\author{A. S. Rodin}
  \affiliation{Centre for Advanced 2D Materials and Graphene Research Centre, National University of Singapore, Singapore 117542, Singapore}

\author{A. Carvalho}
  \affiliation{Centre for Advanced 2D Materials and Graphene Research Centre, National University of Singapore, Singapore 117542, Singapore}

\author{A. H. Castro Neto}
  \affiliation{Centre for Advanced 2D Materials and Graphene Research Centre, National University of Singapore, Singapore 117542, Singapore}

\date{\today}

\begin{abstract}
The relation between unusual Mexican-hat band dispersion, ferromagnetism and ferroelasticity is investigated using a combination of analytical, first-principles and phenomenological methods. The class of material with Mexican-hat band edge is studied using the $\alpha$-SnO monolayer as a prototype. Such band edge causes a van Hove singularity diverging with $\frac{1}{\sqrt{E}}$, and in p-type material leads to spatial and/or time-reversal spontaneous symmetry breaking. We show that an unexpected multiferroic phase is obtained in a range of hole density for which the material presents ferromagnetism and ferroelasticity simultaneously.
\end{abstract}

\maketitle

Recently, a number of two-dimensional (2D) materials have been found to exhibit band dispersions unrelated to their parent three-dimensional (3D) materials. One curious example are the materials with Mexican-hat band which results in van Hove singularities (VHS) with $\frac{1}{\sqrt{E}}$ divergence in the density of states (DOS). This novel material class includes gallium or indium monochalcogenides \cite{jap.118.075101}, and bilayer graphene under electric field \cite{bilayer_graphene,bilayer_graphene_dft,bilayer_graphene_exp}. This leads an electronic instability, often resulting in magnetism, distortion with spatial symmetry breaking, or superconductivity.

First-principles calculations show that GaS \cite{gas_monolayer} and GaSe \cite{gas_monolayer,gase_monolayer} monolayers become ferromagnetic when hole-doped. Here, we demonstrate that this finding is general for materials with Mexican-hat band edges (MHBEs). In addition, we show that ferromagnetism (FM) and ferroelasticity (FE) not only may arise in 2D materials with MHBE, but that both ferroic orders may even be stable simultaneously.

This multiferroicity offers a promising approach to achieve controllable ferromagnetism in 2D materials, where the doping density can be varied by external gating. In fact, most attempts so far to achieve magnetism in 2D have made recourse to magnetic atoms \cite{ramasubramaniam2013mndoped,co-doped_phosphorene}, edges \cite{half-metallic_graphene,phosphorene_nanoribbon} or other defects \cite{liu2014two}. However, these modified materials have been difficult to synthesize and characterize experimentally, and are far from controllable. In contrast, in materials with Mexican-hat like bands, magnetism is inherent. Thus, this class of magnetic materials stands apart within 2D crystals, and may well become the much sought-after monolayer ferromagnetic element needed for designing fully 2D spintronic devices, with the added advantage of allowing for electrical and mechanical tuning of the magnetic state.

Herein, we investigate the relation between Mexican-hat band edges, ferromagnetism and ferroelasticity. As a prototype of this phenomenon, we use p-doped $\alpha$-SnO, in which ferroelastic and ferromagnetic orders coexist.

\begin{figure}[!hbt]
    \centering
        \includegraphics[width=0.49\textwidth]{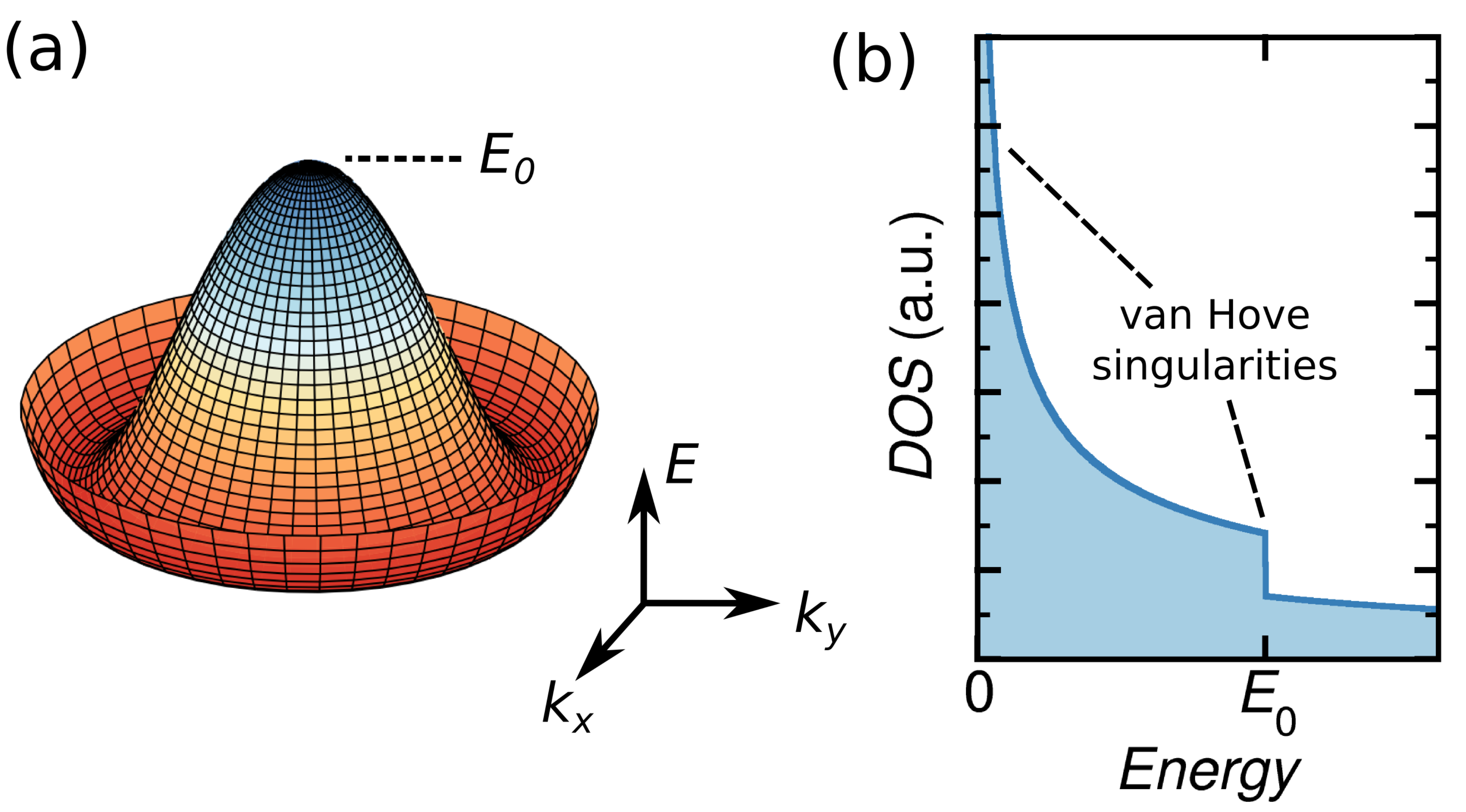}
    \caption{\textbf{Mexican-hat band edge model.} (a) Schematic representation of a MHBE with depth $E_0$. (b) Density of states of two VHS: One at $E=0$ with $\frac{1}{\sqrt{E}}$ divergence, and another with Heaviside step function discontinuity at $E=E_{0}$.}
    \label{fig:Fig1}
\end{figure}

Mexican-hat bands are described by
\begin{equation}
  E(k) = A k^4 + B k^2 + E_{0},
\end{equation}
where $k=|\bf k|$ refers to a point of the reciprocal space. The $A$ and $B$ constants are: $A<0$ and $B>0$ for valence band edge, or $A>0$ and $B<0$ for conduction band edge. The hat depth, $E_0$, is defined as the energy difference between the local band extrema. This kind of band edge is shown schematically in Fig. \ref{fig:Fig1}(a) for $A>0$ and $B<0$. The MHBE have global extrema in an annular region with radius $k_0 = \sqrt{\frac{-B}{2A}}$, and local extrema at $\Gamma$-point. Setting the annular band extrema to zero, the hat depth becomes $E_0 = \frac{B^2}{4A}$. These bands possess two van Hove singularities (VHS): (i) a divergence in the density of states (DOS) with $\frac{1}{\sqrt{E}}$ behaviour at $E=0$, (ii) a Heaviside step function discontinuity at $E=E_0$. Thus, the DOS for MHBE is given by
\begin{equation}
  DOS(E) = \frac{1}{4\pi\sqrt{A}}\frac{1}{\sqrt{E}}\left[ 1 - \frac{\Theta(E-E_0)}{2} \right]
\end{equation}
for the conduction band edge ($A>0$). This DOS is shown in Fig. \ref{fig:Fig1}(b). Details of the calculation of this DOS are found in Supplementary Material (SM) \cite{supp_info}.

We use state-of-art first principles calculations based on density functional theory (DFT) \cite{hohenberg1964,kohn1965} as implemented in the Vienna \textit{Ab Initio} Simulation Package (\textsc{Vasp}) \cite{vasp1,vasp2}. The external potential is given by the projector augmented-wave (PAW) approximation \cite{blochl1994}, and the exchange-correlation functional is given by the generalized gradient approximation parameterized by Perdew--Burke--Ernzerhof (GGA-PBE) \cite{perdew1996generalized} or the hybrid exchange-correlation functional parameterized by Heyd--Scuseria--Ernzerhof (HSE06) \cite{heyd2003hybrid}. The planewave basis with a kinetic energy cutoff of 800~eV is used in all first-principles calculations. The $k$-points samples in the Brillouin zone are calculated with the $\Gamma$-centered Monkhorst--Pack algorithm \cite{monkhorst1976} with $15\times 15\times 1$ grid for monolayers and bilayers, and $15\times 15\times 10$ for bulk geometries. All geometries are relaxed until residual forces are smaller than 10$^{-3}$~eV/\AA, and using nonlocal van der Waals forces in the optB86b-vdW approximation \cite{vdw_1,vdw_2}. We use a vacuum spacing of 20~\AA\ along the $z$-axis to avoid spurious interactions.

\begin{figure}[!htb]
    \centering
        \includegraphics[width=0.49\textwidth]{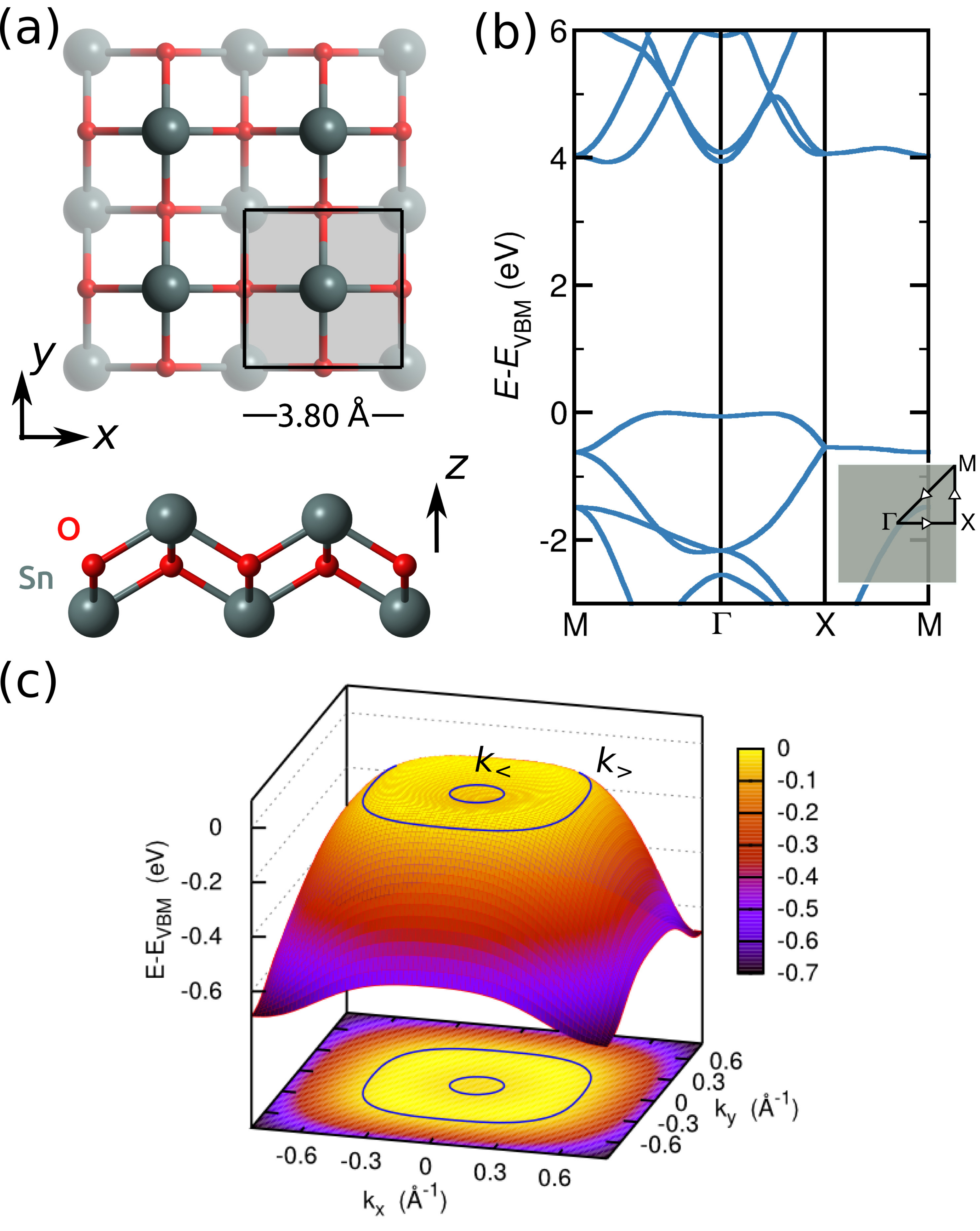}
    \caption{\textbf{$\alpha$-SnO monolayer with tetragonal symmetry.} (a) Ball-and-stick representation of $\alpha$-SnO monolayer from topview (top panel) and sideview (bottom panel). (b) First principles electronic band structure of $\alpha$-SnO monolayer at HSE06 level. (\textbf{c}) Mexican-hat valence band surface at PBE level. Contour lines at $E-E_{VBM} = -0.030$~eV in the valence band are labeled by $k_{<}$ and k$_{>}$.}
    \label{fig:Fig2}
\end{figure}

In order to study the consequences of the DOS divergence for a particular case, we consider monolayer $\alpha$-SnO \cite{sno_1}. The crystal structure of a $\alpha$-SnO monolayer has tetragonal symmetry, see Fig. \ref{fig:Fig2}(a). The oxygen atoms are arranged in a planar square sublattice, while the tin atoms form alternating pyramids with square bases bounded by the oxygen atoms \cite{sno_2,sno_3,sno_4,sno_5}. Using first-principles calculations with van der Waals (vdW) forces, we find a lattice constant of $a_{1L} = 3.803$~\AA\ for $\alpha$-SnO monolayers. This lattice constant is slightly smaller than the one calculated for bilayer and bulk, $a_{2L} = 3.820$~\AA\ and $a_{\rm bulk} = 3.838$~\AA, respectively. Comparisons of lattice constants with different parameterization and experimental values are shown in SM \cite{supp_info}. The $\alpha$-SnO monolayer is an insulator with fundamental indirect bandgap of $3.93$~eV at HSE06 level, as shown in Fig. \ref{fig:Fig2}(b). The optical bandgap is about $4.00$~eV at $\Gamma$-point, with the same exchange-correlation functional. In contrast, the $\alpha$-SnO bulk is a semiconductor with fundamental indirect bandgap of $0.7$~eV and optical bandgap of $2.7$~eV \cite{sno_4,sno_6}. Since the valence band of the monolayer is a shallow MHBE at the $\Gamma$-point, we expect an electronic instability via VHS.

As an approximation, we can consider the valence band maxima as the annular region with radius $k_0$. For Fermi levels between the local band extrema, the Fermi surface is formed by two circular regions with radii $k_{<}$ and $k_{>}$, as shown in Fig. \ref{fig:Fig2}(c). For band structures calculated with HSE06, the Mexican-hat depth is $E_{0}^{\rm HSE06}= -0.052$~eV. With the PBE functional, the depth is $E_{0}^{\rm PBE} = -0.037$~eV. For Fermi levels lower than $E_0$, the Fermi surface is formed by only one ring with radius $k_{>}$. Note that with increasing radii $k_{<}$ and $k_{>}$, there is a tetragonal warping.

\begin{figure*}[!htb]
    \centering
        \includegraphics[width=0.98\textwidth]{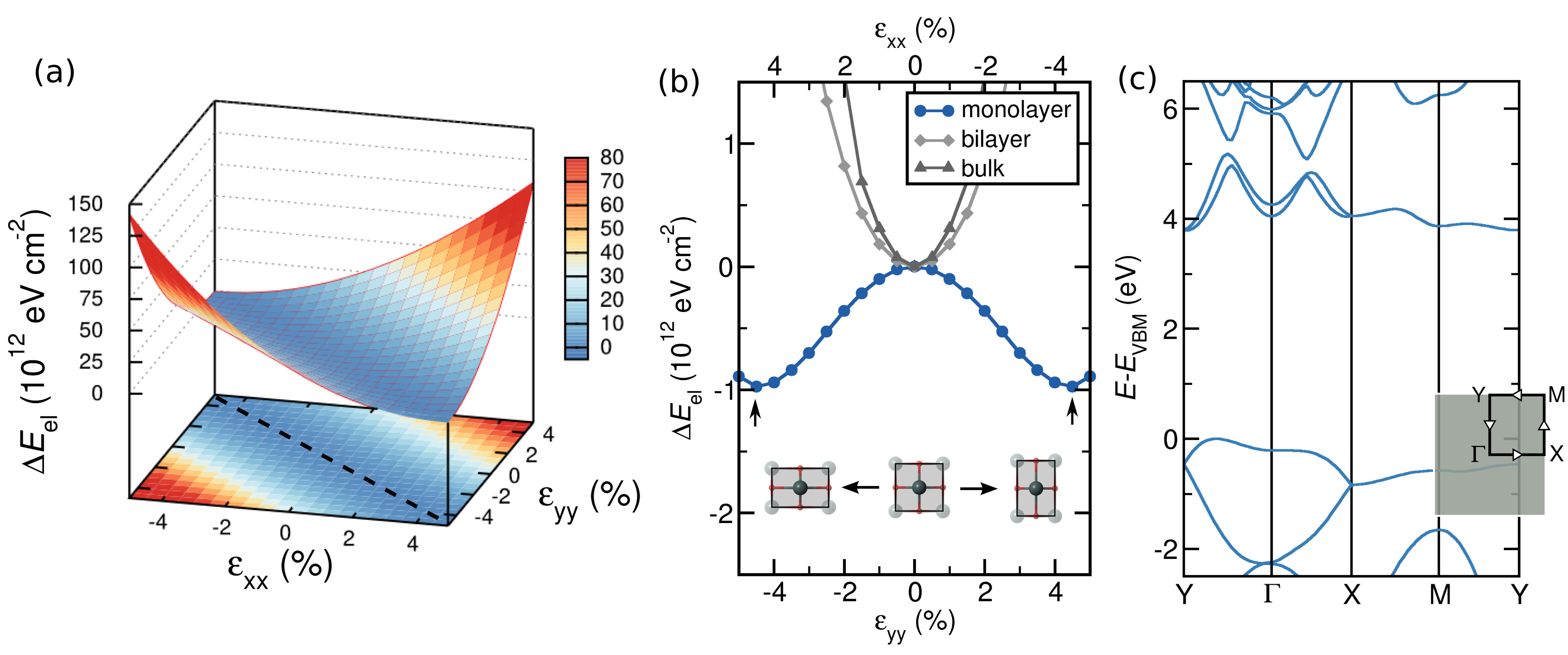}
    \caption{\textbf{Ferroelasticity in $\alpha$-SnO monolayer.} (a) Elastic energy in function of biaxial strain ($\varepsilon_{xx}$ and $\varepsilon_{yy}$) in $\alpha$-SnO monolayer. The antisymmetric diagonal strain ($\varepsilon_{yy} = - \varepsilon_{xx}$) is shown in dashed black line. (b) Elastic energies of $\alpha$-SnO monolayer, bilayer and bulk as function of antisymmetric diagonal strain. Inset: $\alpha$-SnO unit cells with schematic representation of the antisymmetric diagonal strains. (c) Electronic band structure of distorted $\alpha$-SnO with $\varepsilon_{xx} = -\varepsilon_{yy} = 4.5$ \% at HSE06 level.}
    \label{fig:Fig3}
\end{figure*}

Thus, the most interesting situation to consider, aside from the intrinsic material, is p-type material where the Fermi level lies close to the singularity.
This is a realistic scenario, since previous studies of $\alpha$-SnO bulk show that Sn vacancies are the most stable native defects \cite{sno_defects}. These defects are acceptors, doping the $\alpha$-SnO with holes. 

\begin{figure}[!htb]
    \centering
        \includegraphics[width=0.49\textwidth]{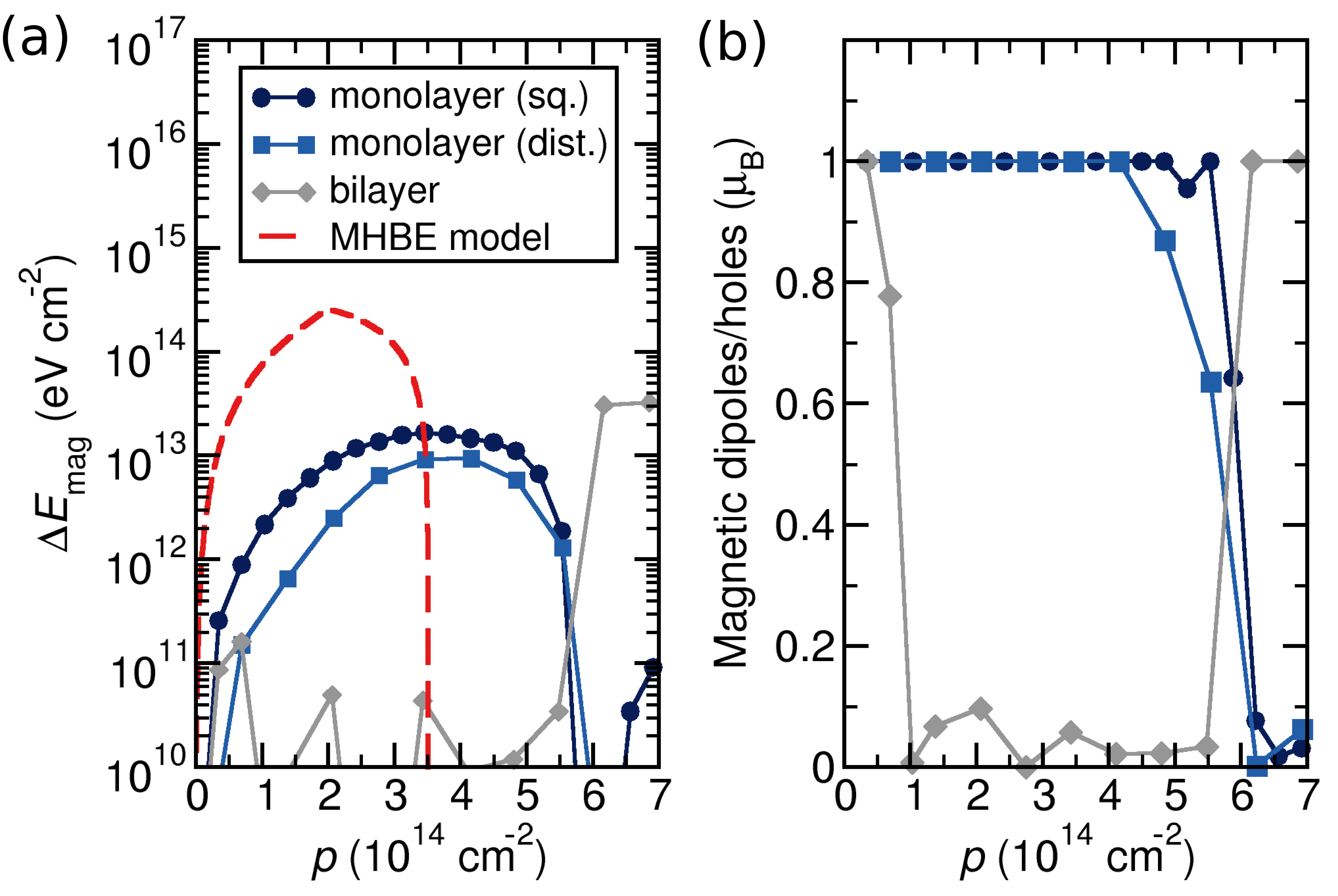}
    \caption{\textbf{Ferromagnetism in $\alpha$-SnO.} (a) Difference of energies between PM and FM orders as a function of hole density. First principles calculations and MHBE model (red). (b) Magnetic dipoles per holes as a function of hole density.}
    \label{fig:Fig4}
\end{figure}

Given the high density of states close to the band edge, we first assess the structural stability of p-type material. The elastic energy $\Delta E_{\rm el}$ is calculated as a function of the strain components $\varepsilon_{xx}$ and $\varepsilon_{yy}$ varying the lattice constants in $x$ and $y$ direction, and relaxing the internal coordinates. The surface $\Delta E_{\rm el}(\varepsilon_{xx}, \varepsilon_{yy})$ shown in Fig. \ref{fig:Fig3}(a) has two degenerate minima in $\left(+4.5~\%, -4.5~\% \right)$ and $\left(-4.5~\%, +4.5~\% \right)$, being characteristic of ferroelasticity. The two minima can be better seen in the transverse section $\varepsilon_{xx} = -\varepsilon_{yy}$ shown in Fig. \ref{fig:Fig3}(b). In contrast, the $\alpha$-SnO bilayer and bulk only have a minimum at $\varepsilon_{xx}=\varepsilon_{yy} = 0$. Thus, only the $\alpha$-SnO monolayer has FE order with intrinsic strain $\varepsilon_{xx}=-\varepsilon_{yy} = \pm 4.5~\%$. This FE order breaks the point symmetry from $C_{4v}$ to $C_{2v}$. At the minimum energy, the band structure of the $\alpha$-SnO monolayer changes to have two valleys (between $\Gamma$ and $Y$, and between $\Gamma$ and $-Y$), see Fig. \ref{fig:Fig3}(c). The DOS of $\alpha$-SnO with $C_{4v}$ symmetry, $C_{2v}$ symmetry, and $\alpha$-SnO bilayer is given in SM \cite{supp_info}.

Other electronic instability driven by the exchange interaction can be observed when we dope the $\alpha$-SnO with holes. Increasing the hole density in monolayers leads to magnetization. This minimizes the magnetic energy, $\Delta E_{\rm mag}$, which we define as the difference between the paramagnetic and ferromagnetic systems. Positive values in Fig. \ref{fig:Fig4}(a) of $\Delta E_{\rm mag}$ indicate the preference for the ferromagnetic order. Both square and distorted monolayers demonstrate magnetization up to $p = 6 \cdot 10^{14}$ cm$^{-2}$, the square monolayers being more stable at $p = 3.46 \cdot 10^{14}$ cm$^{-2}$. The $\alpha$-SnO bilayer is only able to stabilize FM at higher hole densities, when the Fermi level reaches other peaks in the density of states. The magnetic dipole moment that arises in p-doped $\alpha$-SnO monolayer increases linearly with $p$, so that the ratio of the magnetic dipole and holes is close to 1~$\mu_{\rm B}$/holes up to $5\cdot 10^{14}$ cm$^{-2}$, as shown in Fig. \ref{fig:Fig4}(b). In the case $p = 3.46 \cdot 10^{14}$ cm$^{-2}$, we observe that the FM $\alpha$-SnO monolayer is a half-metal phase wherein the spin up peak in the DOS are filled, while the spin down peak are empty, as shown in SM \cite{supp_info}.

In order to show that FM order can be achieved in any 2D material with MHBE, we construct a general model based on the Mexican-hat-like single band. This model can be applied to $\alpha$-SnO monolayers, as well as to other materials with MHBE, such as GaS \cite{gas_monolayer} and GaSe \cite{gase_monolayer} monolayers.

The model is based on the kinetic and exchange energies for systems with ferromagnetic (FM) and paramagnetic (PM) order. From these energies, we can obtain the energy difference between fully spin-polarized and unpolarized systems (see SM \cite{supp_info} for details):
\begin{equation}
  \Delta E_{\rm mag} = p^2 \left\{ \pi^2 Ap -\frac{e^2}{2}\sqrt{\frac{-A}{2B}} \left[ 3 + 2 \ln\left( \frac{-B}{Ap\pi}\right) \right] \right\}.
\end{equation}
The $A$ and $B$ constants can be extracted from DFT, or fitted from experiments. The magnetic energy calculated with this model is also shown in Fig. \ref{fig:Fig4}(a). While the first-principles calculations show a maximal FM stability at $p\approx 3.5 \cdot 10^{14}$ cm$^{-2}$ and FM-PM transition at $p\approx 6 \cdot 10^{14}$ cm$^{-2}$, the effective model show the maximal FM stability at $p \approx 2 \cdot 10^{14}$ cm$^{-2}$ and $p \approx 3.5 \cdot 10^{14}$ cm$^{-2}$.

\begin{figure}[!htb]
    \centering
        \includegraphics[width=0.49\textwidth]{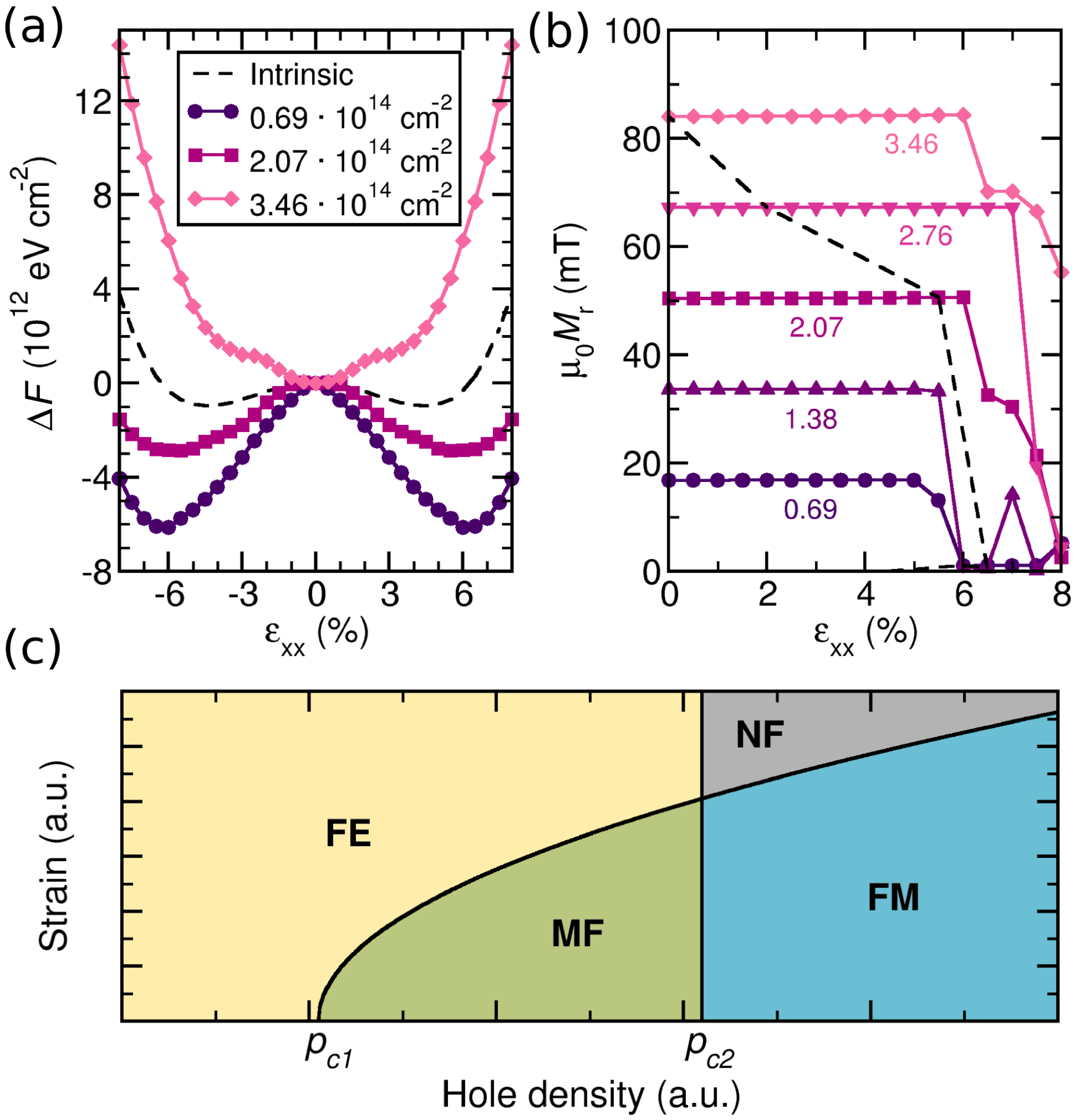}
    \caption{\textbf{Multiferroic phases in $\alpha$-SnO monolayer and models.} (a) Free energy density as a function of the strain $\varepsilon_{xx}$ for intrinsic and p-doped $\alpha$-SnO. (b) Magnetization as a function of the strain $\varepsilon_{xx}$. Each curve is labeled by the hole density in $10^{14}$ cm$^{-2}$. The black dashed line shows the minimum energy strain. (c) Schematic phase diagram based on the generalized Landau model for multiferroicity (ferroelasticity and ferromagnetism) in $\alpha$-SnO monolayer.}
    \label{fig:Fig5}
\end{figure}

In addition to FE and FM orders in the $\alpha$-SnO monolayers, we can also observe phases where both orders are stabilized simultaneously. This regime, designated multiferroic (MF), is usually found in three-dimensional materials with perovskite structures, such as BiFeO$_{3}$ or BaTiO$_{3}$ \cite{perovskite_1,perovskite_2} (with ferroelectricity, instead ferroelasticity). However, so far, MF order has been found for no other 2D material. Varying the antisymmetric diagonal strain ($\varepsilon_{xx} = -\varepsilon_{yy}$) for p-doped $\alpha$-SnO monolayers, we observe that some systems with low hole densities ($0.69 \cdot 10^{14}$ cm$^{-2}$ and $2.07 \cdot 10^{14}$ cm$^{-2}$) reveal FE with intrinsic strains $\varepsilon_{xx} = 6.0$~\% and $\varepsilon_{xx} = 5.5$~\%. For the hole density $p = 3.46 \cdot 10^{14}$ cm$^{-2}$, the system is paraelastic (no nonzero intrinsic strain). The free energy densities for these systems doped with holes are shown in Fig. \ref{fig:Fig5}(a). Meanwhile, calculating the remanent magnetization ($M_{r}$) for these systems as function of strain, we observe that systems doped with holes up to $p = 1.38 \cdot 10^{14}$ cm$^{-2}$ stabilize with FE order, but not with FM. For hole densities from $2.07 \cdot 10^{14}$ cm$^{-2}$ to $2.76 \cdot 10^{14}$ cm$^{-2}$, both FE and FM are stable. Finally, for hole densities greater than $3.46 \cdot 10^{14}$ cm$^{-2}$, only FM orders are stable. The remanent magnetization as functions of strain is shown in Fig. \ref{fig:Fig5}(b) for several values of hole densities. The maximum remanent magnetization found was $\mu_0 M_{r} = 135$~mT at $p= 5.54 \cdot 10^{14}$ cm$^{-2}$. The configurations of magnetization and strain that minimize the energies are shown in Fig. \ref{fig:Fig5}(b).

The phase transitions in these MF 2D materials can be described by a generalized Landau theory in which the free energy density is written as
\begin{equation}
  F = \alpha(p) M^2 + \beta M^4 + \gamma \varepsilon^2 M^2 + \eta(p) \varepsilon^2 + \lambda \varepsilon^4,
\end{equation}
where $\beta$, $\gamma$ and $\lambda$ are positive constants. The functions $\alpha(p)$ and $\eta(p)$ can change the sign within the hole density range, leading to phase transitions. Close to these phase transition, the functions are given by $\alpha(p) = \alpha_0 (p_{c1} - p)$ and $\eta(p) = \eta_0(p - p_{c2})$, where $\alpha_0,\eta_0$ are positive constants, and $p_{c1}$ and $p_{c2}$ are the critical hole densities for FM and FE, respectively. Fixing the magnetization to zero, the phase transition from ferroelastic (FE) to paraelastic (PE) occurs at $p=p_{c2}$. The remaining three terms in free energy involving magnetization, create a phase transition at $\varepsilon_{pt} = \sqrt{\frac{\alpha_0}{\gamma}(p-p_{c1})}$. At higher strains and higher hole densities, the system is nonferroic (NF). Therefore, the physical condition for the multiferroic order to take place in a 2D material is $p_{c1} < p_{c2}$, as shown in Fig. \ref{fig:Fig5}(c). If $p_{c1} > p_{c2}$, the FE and FM will not exist at the same time. For the record, first principles results for $\alpha$-SnO monolayers provide $p_{c1} \approx 2.1 \cdot 10^{14}$ cm$^{-2}$ and $p_{c2} \approx 6.2 \cdot 10^{14}$ cm$^{-2}$, agreeing with the generalized Landau theory.

In conclusion, we show that 2D materials, such as $\alpha$-SnO monolayer, can be ferromagnetic, ferroelastic, or even multiferroic, depending on the hole density. This particular material is the first realization of multiferroic order in two dimensions. This ordering is driven by the $\frac{1}{\sqrt{E}}$ divergence in the density of states. From a single-band model with Mexican-hat dispersion, we demonstrate that the ferromagnetism can be induced in a range of hole densities. Furthermore, we present a generalized Landau model where ferromagnetism, ferroelasticity, and multiferroicity can be stabilized. Using first-principles calculations on $\alpha$-SnO as a prototype, we calculate the hole density range of 2--3 $\cdot 10^{14}$ cm$^{-2}$. However, since the multiferroicity is caused by the Mexican-hat band dispersion, we can infer that other 2D materials can also manifest this order. In other words, a novel class of multiferroic 2D systems can be stabilized, leading to spontaneous symmetry breaking and a nontrivial control of the magnetism and strain.

\begin{acknowledgments}
The authors acknowledge the National Research Foundation, Prime Minister Office, Singapore, under its Medium Sized Centre Programme and CRP award ``\textit{Novel 2D materials with tailored properties: Beyond graphene}'' (R-144-000-295-281). The first-principles calculations were carried out on the Centre for Advanced 2D Materials and Graphene Research Centre high-performance computing facilities.
\end{acknowledgments}


\begin{thebibliography}{50}

\bibitem{jap.118.075101}
D. Wickramaratne, F. Zahid, and R. K. Lake, Electronic and thermoelectric properties of van der Waals materials with ring-shaped valence bands. {\it J. Appl. Phys.} {\bf 118}, 075101 (2015).

\bibitem{gase_monolayer}
T. Cao, Z. Li, and S. G. Louie, Tunable magnetism and half-metallicity in hole-doped monolayer GaSe. {\it Phys. Rev. Lett.} {\bf 114}, 236602 (2015).

\bibitem{gas_monolayer}
S. Wu, X. Dai, H. Yu, H. Fan, J. Hu, and W. Yao, Magnetisms in $p$-type monolayer gallium chalcogenides (GaS, GaSe). arXiv:1409.4733v2

\bibitem{bilayer_graphene}
E. V. Castro, K. S. Novoselov, S. V. Morozov, N. M. R. Peres, J. M. B. L. dos Santos, J. Nilsson, F. Guinea, A. K. Geim, and A. H. Castro Neto, Biased Bilayer Graphene: Semiconductor with a Gap Tunable by the Electric Field Effect, {\it Phys. Rev. Lett.} {\bf 99}, 216802 (2007).

\bibitem{bilayer_graphene_dft}
H. Min, B. Sahu, S. K. Banerjee, and A. H. MacDonald, \textit{Ab initio} theory of gate induced gaps in graphene bilayers, \textit{Phys. Rev. B} {\bf 75}, 155115 (2007).

\bibitem{bilayer_graphene_exp}
Y. Zhang, T.-T. Tang, C. Girit, Z. Hao, M. C. Martin, A. Zettl, M. F. Crommie, Y. R. Shen, and F. Wang, Direct observation of a wide tunable bandgap in bilayer graphene, \textit{Nature} {\bf 459}, 820 (2009).

\bibitem{ramasubramaniam2013mndoped}
A.~Ramasubramaniam and D.~Naveh, Mn-doped monolayer MoS$_2$: An atomically thin dilute magnetic semiconductor, \textit{Phys. Rev. B} {\bf 87}, 195201 (2013).

\bibitem{co-doped_phosphorene}
L. Seixas, A. Carvalho, and A. H. Castro Neto, Atomically thin dilute magnetism in Co-doped phosphorene, {\it Phys. Rev. B} {\bf 91}, 155138 (2015).

\bibitem{half-metallic_graphene}
Y.-W. Son, M. L. Cohen, and S. G. Louie, Half-metallic graphene nanorribons, \textit{Nature} {\bf 444}, 347 (2006).

\bibitem{phosphorene_nanoribbon}
Y. Du, H. Liu, B. Xu, L. Sheng, J. Yin, C.-G. Duan, and X. Wan, Unexpected magnetic semiconductor behavior in zigzag phosphorene nanoribbons driven by half-filled one dimensional band. {\it Sci. Rep.} {\bf 5}, 8921 (2015).

\bibitem{defects_mos2}
P.~Tao, H.~Guo, T.~Yang, Z.~Zhang, Strain-induced magnetism in MoS$_{2}$ monolayer with defects, \textit{J. Appl. Phys.} {\bf 115}, 054305 (2014).

\bibitem{liu2014two}
Y.~Liu, F.~Xu, Z.~Zhang, E.~S.~Penev and B.~I.~Yakobson, Two-Dimensional Mono-Elemental Semiconductor with Electronically Inactive Defects: The Case of Phosphorus, \textit{Nano Lett.} {\bf 14}, 6782 (2014).

\bibitem{supp_info}
Supplementary material is available in the online version of the paper.

\bibitem{hohenberg1964}
P. Hohenberg and W. Kohn, Inhomogeneous Electron Gas, \textit{Phys. Rev.} {\bf 136}, B864 (1964).

\bibitem{kohn1965}
W. Kohn and L. J. Sham, Self-consistent equations including exchange and correlation effects, \textit{Phys. Rev.} {\bf 140}, A1133 (1965).

\bibitem{vasp1}
G.~Kresse and J.~Furthmüller, Effecient itarative schemes for {\it ab initio} total-energy calculations using a plane-wave basis set, \textit{Phys. Rev. B} {\bf 54}, 11169--11186 (1996).

\bibitem{vasp2}
G.~Kresse and D.~Joubert, From ultrasoft pseudopotentials to the projector augmented-wave method, \textit{Phys. Rev. B} {\bf 59}, 1758--1775 (1999).

\bibitem{blochl1994}
P.~E. Blöchl, Projector augmented-wave method, \textit{Phys. Rev. B} \textbf{50}, 17953--17979 (1994).

\bibitem{perdew1996generalized}
J. P. Perdew, K. Burke and M. Ernzerhof, Generalized Gradient Approximation Made Simple, \textit{Phys. Rev. Lett.} {\bf 77}, 3865 (1996).

\bibitem{heyd2003hybrid}
J. Heyd, G. E. Scuseria and M. Ernzerhof, Hybrid functionals based on a screened Coulomb potential, \textit{J. Chem. Phys.} {\bf 118}, 8207 (2003).  {\bf 124}, 219906 (2006)

\bibitem{monkhorst1976}
H.~J.~Monkhorst and J.~D.~Pack, Special points for Brillouin-zone integrations, \textit{Phys. Rev. B} {\bf 13}, 5188 (1976).

\bibitem{vdw_1}
J.~Klime\v{s}, D.~R.~Bowler, and A. Michaelides, Chemical accuracy for the van der Waals density functional, \textit{J. Phys. Condens. Matter} {\bf 22}, 022201 (2010).

\bibitem{vdw_2}
J.~Klime\v{s}, D.~R.~Bowler, and A. Michaelides, Van der Waals density functionals applied to solids, \textit{Phys. Rev. B} {\bf 83}, 195131 (2011).

\bibitem{sno_1}
A.~K.~Singh and R. G. Hennig, Computational prediction of two-dimensional group-IV monochalcogenides, \textit{Appl. Phys. Lett.} {\bf 105}, 042103 (2014).

\bibitem{sno_2}
M.~Meyer, G. Onida, A. Ponchel, L. Reining, Electronic structure of stannous oxide, \textit{Comp. Mat. Sci.} {\bf 10}, 319 (1998).

\bibitem{sno_3}
A.~Walsh and G.~W.~Watson, Electronic structures of rocksalt, litharge, and herzenbergite SnO by density functional theory, \textit{Phys. Rev. B} {\bf 70}, 235114 (2004).

\bibitem{sno_4}
Y.~W.~Li, Y.~Li, T.~Cui, L.~J.~Zhang, Y.~M.~Ma and G.~T.~Zou, The pressure-induced phase transition in SnO: a first-principles study, \textit{J. Phys. Condens. Matter} {\bf 19}, 425230 (2007).

\bibitem{sno_5}
J. P. Allen, D. O. Scanlon, S. C. Parker, G. W. Watson, Tin Monoxide: Structural Prediction from First Principles Calculations with van der Waals Corrections, \textit{J. Phys. Chem. C} {\bf 115}, 19916 (2011).

\bibitem{sno_6}
K. Govaerts, R. Saniz, B. Partoens and D. Lameon, van der Waals bonding and the quasiparticle band structure of SnO from first principles, \textit{Phys. Rev. B} {\bf 87}, 235210 (2013).

\bibitem{sno_defects}
A. Togo, F. Oba, I. Tanaka, K. Tatsumi, First-principles calculations of native defects in tin monoxide, \textit{Phys. Rev. B} {\bf 74}, 195128 (2006).

\bibitem{perovskite_1}
S.-C. Cheong, M. Mostovoy, Multiferroics: a magnetic twist for ferroelectricity, \textit{Nat. Mat.} {\bf 6}, 13 (2007).

\bibitem{perovskite_2}
D. Khomskii, \textit{Physics} {\bf 2}, 20 (2009).

\end{thebibliography}
\end{document}